\documentclass{llncs}
\usepackage{algorithmic}
\usepackage{algorithm}
\usepackage{amssymb}
\usepackage{url}
\usepackage{graphicx}
\usepackage{color}                           

\begin{document}
\mainmatter
\title{Approximate Schedules for Non-Migratory Parallel Jobs in Speed-Scaled Multiprocessor Systems} 
\authorrunning{A. Kononov \and Yu. Kovalenko}
\titlerunning{Approximate Schedules for Non-Migratory Parallel Jobs}
\author{Alexander Kononov\inst{1} \and Yulia Kovalenko\inst{2}} 
\institute{Sobolev Institute of Mathematics,\\
4, Akad. Koptyug avenue, 630090, Novosibirsk, Russia\\
\email{alvenko@math.nsc.ru}\and
Novosibirsk State University,\\
1, Pirogov street, 630090, Novosibirsk, Russia\\
\email{y.kovalenko@nsu.ru}}
\maketitle

\begin{abstract}
We consider a problem of scheduling rigid parallel jobs on variable speed processors so as to minimize the total energy consumption.
Each job is specified by its processing volume and the required number of processors.
We propose new constant factor approximation algorithms for the non-migratory cases when all jobs have a common release time and/or a common deadline.
\end{abstract}
\section{Introduction}
Modern processors can vary their speeds dynamically when jobs are executed.
The instantaneous power required to run a job at speed $s$ is defined by $s^\alpha$, where $\alpha>1$ is a constant.
The energy consumption is the power integrated over time.
We assume that the release times and deadlines are given for jobs
and the aim is to calculate the speeds of the jobs and
to construct a feasible schedule so as to minimize the total energy consumption.
There exist various variants of the speed scaling scheduling, depending on the type of jobs and processors, and system characteristics.
One of the algorithmic and complexity study of this area is devoted to revising classical scheduling problems with dynamic speed scaling (see e.g. \cite{AAG2011,ABKL2012,BG2008,C-A,GHH2016,SSS2015,YDS95} and others).

In this paper we study some particular cases of the
speed scaling scheduling of parallel jobs,
each of them requires several processors simultaneously~\cite{Drozd}.
The motivation to consider parallel jobs consists in the fact that some jobs can not be performed asynchronously on modern computers. Such situation takes place in testing and reliable computing, parallel applications on graphics cards, computer control systems and others.

The energy effective scheduling has been widely investigated for single-proces\-sor jobs (see e.g. reviews \cite{Albers2010,GHH2016}). The preemptive single-processor setting is polynomially solvable.
The algorithms developed in~\cite{LYYu2014,SSS2015} have running time $O(n^2)$.
However, the same problem without preemptions is NP-hard~\cite{AH2012} even in the case of tree structured jobs.
Antoniadis et al.~\cite{AH2012} proposed an $2^{4\alpha-3}$-approximation algorithm for this case and
an $2^{5\alpha-4}$-approximation algorithm for the general non-preemptive case on one processor.
Later, Bampis et al.~\cite{BKLLS2013} presented an
algorithm that achieves $2^{\alpha-1}(1+\varepsilon)^{\alpha}\tilde{B}^{\alpha}$-approximation for the latter problem, where $\tilde{B}_{\alpha}$ is the generalized Bell number.

For the migratory problem, where $m$ parallel processors are available and all jobs are of the
single-processor type, polynomial time algorithms  were given in~\cite{AAG2011,ABKL2012,BG2008,SSS2015}.
As far as we know, the algorithm~\cite{SSS2015} has the best time complexity $O(n^3)$ among the algorithms above.

The preemptive  multiprocessor setting without migration is NP-hard in the strong sense
as proved in~\cite{AMS2007}.
Albers et al.~\cite{AMS2007} provided $\left(\alpha^{\alpha} 2^{4\alpha}\right)$-approximation algorithms
for jobs with unit works and arbitrary deadlines, or arbitrary works and agreeable deadlines,
and an $2\left(2 - \frac{1}{m}\right)^{\alpha}$-approximation algorithm
for jobs with common release times, or common deadlines.
Moreover, they showed that the problem with unit works is polynomially solvable
for agreeable deadlines.
Chen et al.~\cite{CHCYPK2004} proposed  a greedy algorithm with an $1.13$ approximation guarantee for the case when all jobs must be executed in one time interval.
In~\cite{BKLLN2015}, for  non-preemptive instances an $m^{\alpha} \left(\sqrt[m]{n}\right)^{\alpha-1}$-approximation algorithm has been presented,
that explore the idea of transforming an optimal preemptive schedule to a non-preemptive one.

Scheduling of multiprocessor
jobs has been extensively investigated in the case of regular time criteria (see e.g. the book of Drozdowski~\cite{Drozd}), but for the criterion of energy minimization
it is poorly  studied.
Recently, an approximation algorithm  has been proposed in~\cite{KonKov2016}
for the speed scaling scheduling of rigid parallel jobs  with  migration.
The algorithm  returns a solution within
an additive error~$\varepsilon>0$ and runs in time polynomial in $m,$ $1/\varepsilon$  and the input size.
Note that this algorithm is pseudopolynomial and it is based on solving a linear configuration program using the Ellipsoid method. 
In~\cite{KonKov2017}, we developed a strongly
polynomial algorithm that achieves $\left(2-\frac{1}{m}\right)^{\alpha-1}$-approximation ratio for the same problem. 
As well we showed that most of the NP-hardness proofs for scheduling problems with the
maximum lateness criterion  may be easily transformed to their speed scaling counterparts~\cite{KonKov2016}.

\section{Problem Statement and Our Results}
We assume that a computer system consists of $m$ parallel identical processors, which can dynamically change the speed.
A set~$\mathcal{J}=\{1,\dots,n\}$ of parallel jobs is given.
Each job~$j\in \mathcal{J}$ has
a release time $r_j$, a deadline $d_j$ and  a processing volume (work) $W_j$.
The number of processors simultaneously required by job~$j$ is called {\em job size} and denoted by~$size_j$.
Any subset of parallel processors of the given size can be used to execute job $j$.
Jobs with such property are called {\em rigid jobs}~\cite{Drozd}.
Migration of a job $j$ among different subsets of $size_j$ {processors} is disallowed.
Job preemption might or might not be allowed in the exploring of scheduling in this paper.

The speed of a job $j$ is the rate at which its work is completed.
A continuous spectrum of processor speeds is available.
The power consumed when running at speed $s$ is
$s^\alpha$, where $\alpha$ is a constant
close to $3$~\cite{GHH2016}. The energy used is power integrated over time.
Each of $m$ processors may operate at variable speed, but if processors execute the same job simultaneously then all these processors run at the same speed.

A schedule is required to be feasible in the sense that each processor executes at most one job at the time and each
job is processed in the required work between its release time and deadline.
The problem is to find a feasible schedule that minimizes the total energy consumed  on all the processors.
The preemptive and nonpreemptive variants of non-migration  energy-efficient scheduling of rigid jobs~\cite{BKLLS2013,Drozd}
are denoted by  $P|size_j,pmtn*,r_j,d_j|E$ and  ${P|size_j,r_j,d_j|E}$, respectively.

Let us assume that all jobs have a common release time~$r$ and/or a common deadline~$d$.
We present strongly polynomial time algorithms achieving constant factor approximation guarantees for these particular cases.
Our algorithms consist of two stages. At the first stage
we obtain a lower bound on the minimal energy consumption and calculate intermediate execution times of jobs. 
Then, at the second stage, we determine final speeds of jobs and schedule them.

A lower bound on the objective function and intermediate processing times of jobs can be found in $O(n^3)$ time using the method developed in~\cite{KonKov2017}.
The method is based on a reduction of the speed scaling problem to the special min-cost max-flow problem~\cite{SSS2015}. Here we propose more effective approaches for the considered problem instances.

Problems $P|size_j,pmtn*,r_j=r,d_j=d|E$ and $P|size_j,r_j=r,d_j=d|E$
are strongly NP-hard even in the case of single-processor jobs with arbitrary processing volumes as proved in~\cite{AMS2007}.
Simple reductions from 2-PARTITION and 3-PARTITION imply
that  problem $P|size_j,pmtn*,r_j=r,d_j=d|E$ is ordinary NP-hard and problem~$P|size_j,r_j=r,d_j=d|E$ is NP-hard in the strong sense even
if all jobs have unit processing volumes. Using the approach from~\cite{KonKov2016} (Section 3. NP-Hardness Results),
we can show that problem $P|size_j,r_j,=r,d_j|E$ is strongly NP-hard even in the case of two processors.
The proof is almost a step by step reproduction of the NP-hardness proof for $P2|size_j, d_j |L_{\max}$ (see Theorem Al in~\cite{LeeCai1999}).

The paper is organized as follows. In Section~\ref{sec:EqualRandD} we propose an  $\left(2-\frac{1}{m}\right)^{\alpha-1}$-approxi\-mation algorithm
for the non-preemptive scheduling where all jobs are available in one time interval. In Section~\ref{sec:EqualRorD},
 $\left(3-\frac{4}{m+1}\right)^{\alpha-1}$-approximate schedules are
 constructed for the preemptive and non-preemptive problems with jobs sharing a common release time (or symmetrically, a common deadline). 
The last section contains the concluding remarks.

\section{Common Release Date and Deadline}
\label{sec:EqualRandD}
In this section we consider the non-preemptive case of the problem where all jobs arrive at time $r=0$ and have a shared global deadline~$d$.

\textbf{The first stage.}
Now we find auxiliary durations of jobs and a lower bound on the objective function in $O(n\log n)$ time.
Let $m'$ denote the current number of unoccupied processors and $\mathcal{J}'$ be the set of currently considered jobs.
Initially $\mathcal{J}':=\mathcal{J}$ and $m':=m$.

We enumerate the jobs one by one in order of non-increasing  works. If the current job~$i$ has ${W_i\ge\frac{\sum_{j\in \mathcal{J}'} W_j size_j}{m'}}$,
then we assign duration $p_i:=d$ for this job, and set $\mathcal{J}':=\mathcal{J}'\setminus \{i\}$ and $m':=m'-size_i$. After that we go to the next job.
Otherwise, all jobs ${l\in \mathcal{J}'}$ satisfy the inequality  ${W_l<\frac{\sum_{j\in \mathcal{J}'} W_j size_j}{m'}}$,
and we assign durations $p_l:=\frac{W_l m' d}{\sum_{j\in \mathcal{J}'} W_j size_j}$~for~them.

The presented approach guarantees 
that  $p_j \le d$, $\sum_{j\in \mathcal{J}}p_j size_j= md$,
and gives the lower bound on the objective function equal to  $\sum_{j\in \mathcal{J}}size_j W_j^{\alpha} p_j^{1-\alpha}$.
 At the second stage we use the ``non-preemptive list-scheduling'' algorithm~\cite{NS2002} to construct a feasible schedule in interval~$[0,d)$.

\textbf{The second stage.}
Whenever a subset of processors
falls idle, the ``non-preemptive list-scheduling'' algorithm schedules a job that does not require more processors than are available,
until all jobs in $\mathcal{J}$ are assigned. The time complexity of the algorithm is $O(n^2)$.


We claim that the length of the constructed schedule is at most ${\left(2-\frac{1}{m}\right)d}$ (see Lemma~\ref{AppAlgRandD} below).
By increasing the speed of each job in $\left(2-\frac{1}{m}\right)$ times we obtain a schedule of the  length at most $d$.
The total energy consumption  is  increased by a factor $\left(2-\frac{1}{m}\right)^{\alpha-1}$ in comparison with the lower bound.
As a result, we have

\begin{theorem}
A $\left(2-\frac{1}{m}\right)^{\alpha-1}$-approximate schedule
can be found in $O(n^2)$ time for problems $P|size_j,pmtn*,r_j=r,d_j=d|E$ and $P|size_j,r_j=r,d_j=d|E$.
\end{theorem}

Using the results from~\cite{Jh2006}, we conclude that the approximation ratio of $\left(2-\frac{1}{m}\right)$
for the ``non-preemptive list-scheduling'' algorithm is tight even if all jobs have single-processor type.
As a result, the energy consumption is increased in $\left(2-\frac{1}{m}\right)^{\alpha-1}$
times when we put the resulting schedule inside the interval $[0,d)$. Therefore, the approximation ratio of our algorithm is also tight.

\begin{lemma}\label{AppAlgRandD}
Given $m$ processors, an interval $[0,d)$, and a set of jobs $\mathcal{J}$ with processing times
$p_j\le d$ and sizes $size_j$,
where $\sum_{j\in \mathcal{J}}p_j size_j\le md$. The length of the schedule~$S$ constructed
by the ``non-preemptive list-scheduling'' algorithm  is at most $\left(2-\frac{1}{m}\right)d$.
\end{lemma}

\textbf{Proof.} Let $T$ denote the length of schedule~$S$. 
If at least $\frac{m+1}{2}$ processors are used at any time instance in $S$, we have

$$d\ge \frac{1}{m}\sum_{j\in \mathcal{J}}p_j size_j \ge \frac{m+1}{2m} T \ge \frac{T}{2-1/m}.$$

Otherwise, assume that $I$ is the last time interval of schedule~$S$ with $m_I < \frac{m+1}{2}$ processors being used during $I$.
By the construction of $S$ there is a job $j$ that is performed during the whole interval~$I$.
Let $C_j$ be the completion time of $j$ in~$S$.
It is easy to see that at every point in time during interval $[0, C_j-p_j)$ schedule $S$ uses at least $m-size_j+1\ge m-m_I+1$ processors
(otherwise job $j$ should be started earlier).
Moreover, at least $m_I$ processors are utilized in interval $[C_j-p_j,C_j)$, therefore,
each job executed in interval $[C_j, T)$ requires  no less than $m-m_I+1$ processors.
Thus, the total load of all processors $\sum_{j\in \mathcal{J}} p_j size_j$  is at least  ${m_I p_j +(T-p_j)(m-m_I+1)\le dm}$.
If $p_j\ge \frac{T}{2-\frac{1}{m-m_I+1}}$, then $\frac{T}{2-\frac{1}{m}} \le \frac{T}{2-\frac{1}{m-m_I+1}} \le p_j \le d$.

Otherwise as $1\le m_I\le \frac{m}{2} < \frac{m+1}{2}$ we have

$$d\ge \frac{1}{m}\left((m-m_I+1)T - p_j(m-2m_I+1)\right) \ge \frac{T}{m} \left(m-m_I+1- \frac{m-2m_I+1}{2-\frac{1}{m-m_I+1}}\right)$$
$$=\frac{T(m-m_I+1)}{(2m-2m_I+1)}\ge \frac{mT}{2m-1}=\frac{T}{2-\frac{1}{m}}.$$
\qed

\section{Common Release Date or Deadline}
\label{sec:EqualRorD}
In this section we study the problem without migration where all jobs are released at time $r=0$ but have individual deadlines.

\subsection{Preemptive Problem}
\label{PreemptiveRorD}
Here we consider the case when the preemption of jobs is allowed.

\textbf{The first stage.}
It can easily be checked that the optimal energy consumption of rigid jobs with sizes $size_j$ and works $W_j$ on $m$ processors
is at least $\frac{1}{m^{\alpha-1}}$ times that of single-processor jobs with works $size_jW_j$ on one processor
(the result is proved  similarly to  Lemma~1 in~\cite{AMS2007}).
So, if we find an optimal solution of the latter problem, and decrease the speeds of jobs in $m$ times,
then a lower bound on the energy consumption of rigid jobs is obtained.
However such approach may lead to the execution time of a job~$i$ greater than $d_i$.

In~\cite{YDS95}, Yao et al. showed that the preemptive problem on a single processor is solvable in polynomial time.
They proposed an efficient algorithm called \textit{YSD} that repeatedly identifies time intervals of highest density. The density
of an interval $I$ is the total work released and to be completed in $I$ divided by the length of $I$. The algorithm
repeatedly schedules jobs in highest density intervals and takes care of reduced subproblems.

We propose a modification of the algorithm YSD~\cite{YDS95} to obtaining
a lower bound on the minimal energy consumption for the considered problem. At the same time,
for each rigid job $i\in \mathcal{J}$ we find an intermediate duration $p_i$ such that

\begin{equation}p_i\le d_i,\label{eachjob}\end{equation}
\begin{equation}\sum_{j\in \mathcal{J}_i} p_j size_j\le md_i,\ \mathcal{J}_i=\{j\in \mathcal{J}:\ d_j\le d_i\}.\label{eachdeadline}\end{equation}

Now we construct a special schedule for jobs of works $size_jW_j$ on one processor,
which will assure conditions (\ref{eachjob}) and (\ref{eachdeadline}) for the corresponding rigid jobs.
Let $J(t,t'):=\{j\in \mathcal{J}:\ t\le r_j<d_j\le t'\}$ be the jobs which need to be processed in some time interval $[t,t')$.
Initially we have intervals of the form $[0,d_j),\ j\in \mathcal{J},$ as all jobs release at time~$r=0$.\\

\textbf{{\em Modified YSD Algorithm}}\\

\textbf{Step~1.}  Repeat steps 1.1 and 1.2 until  $\mathcal{J}= \emptyset$:\\

\textbf{Step~1.1.} Let $[t,t')$ be the interval with maximum density, i.e., that maximizes

$$\frac{\sum\limits_{j\in J(t,t')} size_jW_j}{t'-t}.$$

\textbf{Step~1.2.} If the inequality

\begin{equation} \frac{W_i}{d_i}\le \frac{\sum_{j\in J(t,t')} size_jW_j}{(t'-t)m} \label{feasibility}\end{equation}
holds for all $i\in J(t,t')$,  then process these jobs in interval $[t,t')$ with speed equal to the maximum density,
i.e., set processing time $p_i^1:=\frac{size_iW_i(t'-t)}{\sum_{j\in J(t,t')} size_jW_j}$ for each job $i\in J(t,t')$.
Then remove the jobs $J(t,t')$ from $\mathcal{J}$, and  adjust the remaining jobs
as if the time interval $[t,t')$ does not exist, i.e., set $r_j:=t'$ for each job $j\in \mathcal{J}$.
Endpoints and densities of intervals are updated for Step~1.1.

Otherwise enumerate all jobs from $J(t,t')$ for which inequality~(\ref{feasibility})
is violated. The current job~$i$ is assigned in interval $[d_i-\frac{d_isize_i}{m},d_i)$, 
i.e., $p_i^1:=\frac{d_isize_i}{m}$. Then remove job~$i$ from $\mathcal{J}$ and delete interval $[d_i-p_i^1,d_i)$ from the further consideration,
i.e., set $d_j:=d_j-p_i^1$ for $d_j\ge d_i$ and $d_j:=d_i-p_i^1$ for $d_i-p_i^1\le d_j< d_i$.

\textbf{Step~2.}  Return the resulting durations $p_i^1$ of jobs $j\in \mathcal{J}$.\\

At least one job is deleted at each call of Step~1.2.
Removing a job requires $O(n)$ additional operations to update information about the remaining jobs and intervals.
Therefore, the running time of the algorithm is $O(n^2)$.

The intermediate processing times~$p_i$ of rigid jobs $i\in \mathcal{J}$ are calculated as $\frac{p_i^1m}{size_i}$,
and give the lower bound $\sum_{j\in \mathcal{J}}size_j W_j^{\alpha} p_j^{1-\alpha}$ on the energy consumption.
The conditions (\ref{eachjob}) and (\ref{eachdeadline}) hold for the computed processing times.

%

At the second stage we use the ``preemptive earliest deadline list-scheduling'' algorithm to construct
a feasible schedule of problem~$P|size_j,pmtn*,r_j=0,d_j|E$. 

\textbf{The second stage.}
The ``preemptive earliest deadline list-scheduling'' algorithm schedules jobs  in order
of non-decreasing deadlines as follows.  If $size_i>\frac{m}{2}$, then job~$i$ is assigned at the end of the current schedule.
Otherwise we start job~$i$ at the earliest time instant when $size_i$ processors are idle and process it during $p_i$ time,
ignoring intervals of jobs with $size_j>\frac{m}{2}$.  The time complexity of the algorithm is $O(n^2)$.

We claim that the completion time~$C_j$ of each job~$j$ in the constructed schedule is at most ${\left(3-\frac{4}{m+1}\right)d_j}$
(see Lemma~\ref{AppAlgRorDpr} below).
Hence an increasing of the speeds  in ${\left(3-\frac{4}{m+1}\right)}$ times yields a feasible schedule.
The total energy consumption  is  increased by a factor $\left(3-\frac{4}{m+1}\right)^{\alpha-1}$. 

Obviously, by interchanging release times and deadlines, the presented algorithm can also handle to the case of jobs with
individual release times but a common deadline.
As a result, we have

\begin{theorem}
A $\left(3-\frac{4}{m+1}\right)^{\alpha-1}$-approximate schedule
can be found in $O(n^2)$ time for problems $P|size_j,pmtn*,r_j=r,d_j|E$ and $P|size_j,pmtn*,r_j,d_j=d|E$.
\end{theorem}

%

\begin{lemma}\label{AppAlgRorDpr}
Given $m$ processors and a set of jobs $\mathcal{J}$ with deadlines~$d_i,$  processing times
$p_i\le d_i$ and sizes $size_i$,
where $\sum_{j\in \mathcal{J}_i}p_j size_j\le md_i$ for each $i\in \mathcal{J}$ with $\mathcal{J}_i=\{j\in \mathcal{J}:\ d_j\le d_i\}$.
The completion time~$C_i$  is at most ${\left(3-\frac{4}{m+1}\right)d_i}$ for each job~$i\in \mathcal{J}$ in the schedule~$S$ constructed
by the ``preemptive earliest deadline list-scheduling'' algorithm.
\end{lemma}

\textbf{Proof.} We consider an arbitrary deadline $d_i$,  where job~$i$ has the maximal completion time $C_i$ in schedule~$S$
among all jobs with deadline equal to $d_i$.

Note that $C_j\le C_i$ for all $j\in \mathcal{J}_i$. Let $S_i$ denote the part of schedule $S$, which
contains only jobs from $\mathcal{J}_i$ and occupies interval $[0,C_i)$.
We will show that $C_i\le {\left(3-\frac{4}{m+1}\right)d_i}$.

If at least $\frac{m+1}{2}$ processors are used at any
time instance in subschedule $S_i$, we have

$$d_i\ge \frac{1}{m}\sum_{j\in \mathcal{J}_i}p_j size_j \ge \frac{m+1}{2m} C_i \ge \frac{C_i}{2-\frac{1}{m}}.$$

Otherwise assume that $l$ is the last job in subschedule~$S_i$, that requires $size_l \le \frac{m}{2}$ processors.
It is easy to see that all time-slots in intervals $[0,C_l-p_l)$ and $[C_l,C_i)$ use
at least $\frac{m+1}{2}$ processors,
 and at least $size_l$ processors are utilized in interval $[C_l-p_l;C_l)$.
Therefore, the total load of all processors in subschedule $S_i$  is at least

$$\frac{m+1}{2}\left(C_i-p_l\right)+size_lp_l\le \sum_{j\in \mathcal{J}_i}p_jsize_j\le d_im.$$

If $p_l\ge \frac{C_i}{3-\frac{4}{m+1}}$, then $C_i \le \left(3-\frac{4}{m+1}\right)d_l \le \left(3-\frac{4}{m+1}\right)d_i$.

Otherwise as $size_l\ge 1$ we have

$$md_i\ge \frac{m+1}{2}C_i - p_l\left(\frac{m+1}{2}-size_l\right) \ge  \frac{m+1}{2}C_i - p_l\frac{m-1}{2}\ge$$
$$\ge C_i \left(\frac{m+1}{2}- \frac{m-1}{2  \left(3-\frac{4}{m+1}\right)} \right)=C_i\frac{(m+1)(3m-1)-(m-1)(m+1)}{2(3m-1)}=$$
$$=C_i\frac{m(m+1)}{3m-1}=\frac{mC_i}{3-\frac{4}{m+1}}$$
\qed

\subsection{Non-Preemptive Problem}
In this subsection we consider the case when preemptions are disallowed and each job requires no more than  $\frac{m}{2}$ processors.

A lower bound on the objective function and intermediate processing times of jobs are calculated as in Subsection~\ref{PreemptiveRorD}. 
But at the second stage we use the ``non-preemptive earliest deadline list-scheduling'' algorithm to construct a schedule.
This algorithm assigns jobs in the schedule as soon as possible in order
of non-decreasing deadlines, and it's time complexity is also $O(n^2)$.

The property of Lemma~\ref{AppAlgRorDpr} holds as well in the presence of the ``non-preemptive earliest deadline list-scheduling'' algorithm
for problem $P|size_j,r_j=0,d_j|E$ with $size_j\leqslant \frac{m}{2}$.
So, an increasing of the job speeds in ${\left(3-\frac{4}{m+1}\right)}$ times in the constructed schedule leads to a feasible solution. 
Thus the following theorem holds.

\begin{theorem}
A $\left(3-\frac{4}{m+1}\right)^{\alpha-1}$-approximate schedule
can be found  in $O(n^2)$ time for problems $P|size_j,r_j=r,d_j|E$  and $P|size_j,r_j,d_j=d|E$ with $size_j\leqslant \frac{m}{2}$.
\end{theorem}

\section{Conclusion}

We have studied the energy minimization under a global release time or a global deadline constraint.
Strongly polynomial time approximation algorithms are developed for the rigid jobs with no migration.
Our algorithms have constant factor approximation guarantees.

Further research might address the approaches to the problems with more
complex structure, where processors are heterogeneous and  jobs have
alternative execution modes with different characteristics.

%
%
%
%





\end{document}